\documentclass[conference]{IEEEtran}

\usepackage[pdftex]{graphicx}
\usepackage{amsmath}
\usepackage{algpseudocode}
\algrenewcommand\algorithmicrequire{\textbf{Input:}}
\algrenewcommand\algorithmicensure{\textbf{Output:}}
\usepackage{soul}
\setul{0.25ex}{}
\usepackage{booktabs}
\usepackage{pifont}
\usepackage{xcolor}
\usepackage{tikz}
\usepackage{url}
\usepackage[hidelinks]{hyperref}

\begin{document}

\title{\LARGE LightningSimV2: Faster and Scalable Simulation for High-Level Synthesis via Graph Compilation and Optimization}

\author{\IEEEauthorblockN{Rishov Sarkar, Rachel Paul, Cong (Callie) Hao}
\IEEEauthorblockA{School of Electrical and Computer Engineering, Georgia Institute of Technology\\
\{\href{mailto:rishov.sarkar@gatech.edu}{\nolinkurl{rishov.sarkar}}, \href{mailto:rachel.paul@gatech.edu}{\nolinkurl{rachel.paul}}, \href{mailto:callie.hao@gatech.edu}{\nolinkurl{callie.hao}}\}\nolinkurl{@gatech.edu}}}

\maketitle

\begin{abstract}
High-Level Synthesis (HLS) enables rapid prototyping of complex hardware designs by translating C or C++ code to low-level RTL code. However, the testing and evaluation of HLS designs still typically rely on slow RTL-level simulators that can take hours to provide feedback, especially for complex designs. A recent work, LightningSim, helps to solve this problem by providing a simulation workflow one to two orders of magnitude faster than RTL simulation. However, it still exhibits inefficiencies due to several types of redundant computation, making it slow for large design simulation and design space exploration. Addressing these inefficiencies, we introduce LightningSimV2, a much faster and scalable simulation tool. LightningSimV2 features three main innovations. First, we perform compile-time static analysis, exploiting the repetitive structures in HLS designs, e.g., loops, to reduce the simulation workload. Second, we 
propose a novel graph-based simulation approach, with decoupled simulation graph construction step and graph traversal step, significantly reducing repeated computation.
Third, benefiting from the decoupled approach, LightningSimV2 can perform incremental stall analysis extremely fast, enabling highly efficient design space exploration of large numbers of complex hardware parameters, e.g., optimal FIFO depths.
Moreover, the DSE is well-suited for parallel computing, further improving the DSE efficiency.
Compared with LightningSim, LightningSimV2 achieves up to 3.5\texttimes{} speedup in full simulation and up to 577\texttimes{} speed up for incremental DSE. Our code is open-source on GitHub at \url{https://github.com/sharc-lab/LightningSim/tree/v0.2.0}.

\end{abstract}

\IEEEpeerreviewmaketitle

\section{Introduction}

\begin{figure*}
    \centering
    \includegraphics[width=0.95\linewidth]{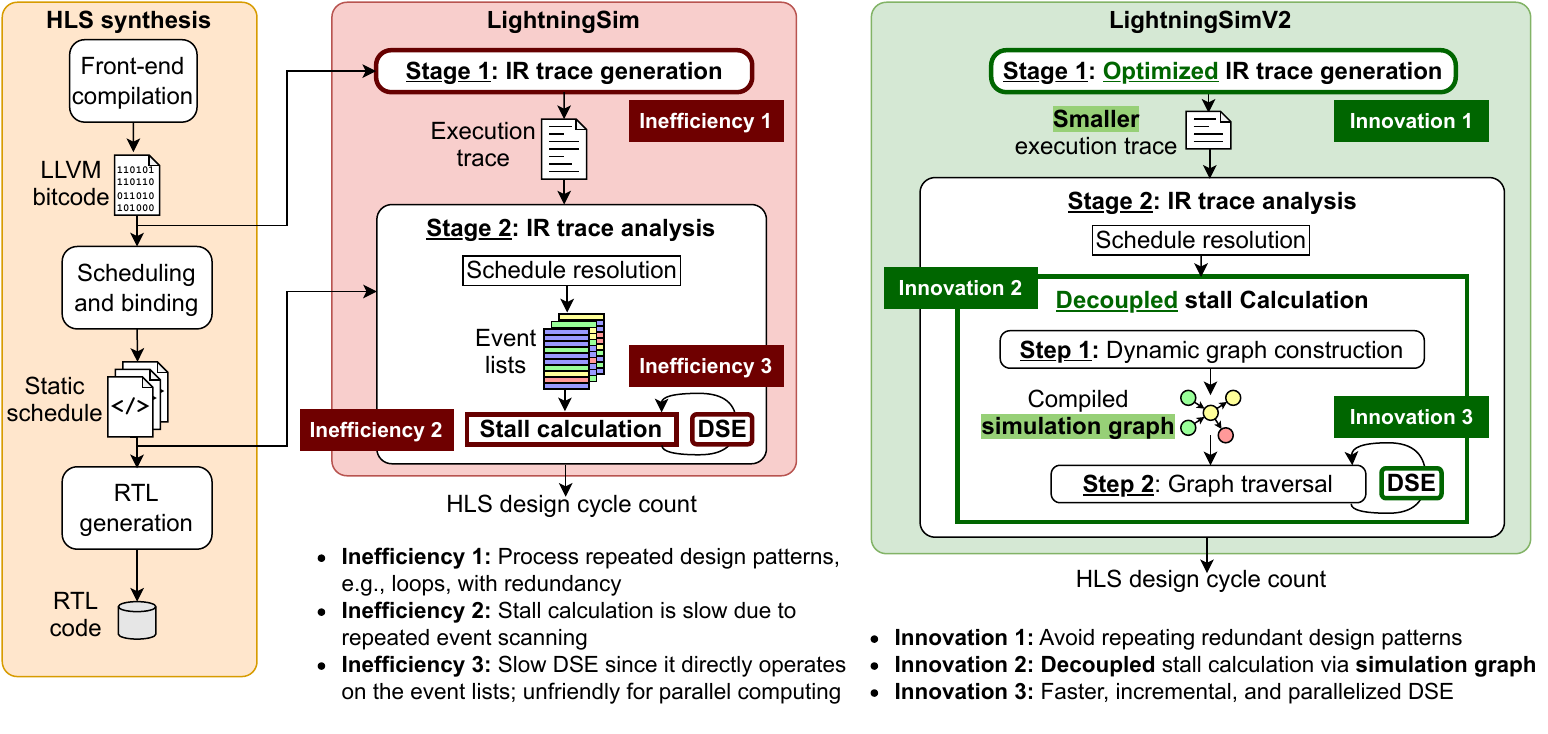}
    \caption{LightningSimV2 addresses three major limitations in LightningSim with three innovations. First, it uses static analysis to reduce the size of the generated execution trace by avoiding repeating redundant design patterns, e.g., loops. Second, it largely speeds up stall calculation by a decoupled graph compilation and graph traversal during trace analysis. Third, it largely speeds up DSE for hardware parameters, e.g., FIFO depths, by only executing the lightweight graph traversal step, which is suitable for scalable and parallel computing with little memory overhead.}
    \label{fig:lightningsimv2}
\end{figure*}

Growing demands for real-time, energy-efficient, high-throughput compute combined with the end of Moore's law have emphasized the need for customized computing architectures more than ever before. Faced with the challenge of creating ever more complex designs, many hardware designers have turned to High-Level Synthesis (HLS), which facilitates the process of hardware coding by raising the level of abstraction from RTL implementation to behavioral modeling in higher-level software languages such as C, C++, or SystemC. HLS tools aim to synthesize this high-level software code, typically annotated with compiler directives to guide the hardware implementation, into RTL code for the same functionality.

HLS greatly improves designer productivity in three ways. First, HLS code can provide a concise specification of hardware structures that would otherwise take much longer to write and test. Second, HLS enables the use of C simulation, in which HLS code is compiled and run using the standard toolchains of the high-level language as a fast and easy way to perform a partial check to ensure that an HLS design functions as intended. Third, the HLS synthesis process provides feedback in the form of latency and resource utilization estimates.

However, in many cases, designers require more precise evaluation of their HLS designs than what is provided by C simulation or the synthesis process. For example, when using the HLS dataflow feature to generate multiple hardware modules communicating via first-in-first-out (FIFO) streams, neither C simulation nor HLS synthesis can detect deadlock conditions caused by FIFO communication. Additionally, HLS synthesis latency estimates become much less reliable as they cannot account for dynamic FIFO communication behavior.

In such cases, where the dynamic behavior of a design limits visibility into its hardware functionality and performance, designers usually resort to slow simulation of the HLS-generated RTL code. This can take hours, especially for complex designs. A recent work, \textbf{LightningSim}~\cite{lightningsim}, provides a much faster alternative to RTL simulation for dynamic behavior modeling and cycle-level estimation of designs built for AMD Vitis HLS. 
As Fig.~\ref{fig:lightningsimv2} shows, LightningSim exploits LLVM intermediate representation (IR) in the synthesis process. It 
has two stages: \textit{IR trace generation} and \textit{IR trace analysis}.
The trace analysis stage further contains a schedule resolution step and a stall calculation step to obtain the total cycle count of an HLS design. %
Furthermore, it features a unique ability to perform 
\textit{incremental stall calculation}
when only certain hardware parameters, namely FIFO buffer depths, are changed. Such changes do not affect HLS scheduling or the execution trace, so only the stall calculation step needs to be re-executed.
LightningSim produces 99.9\%-accurate latency estimates up to 95\texttimes{} faster than RTL simulation in Vitis HLS. 

Despite the speed up over RTL simulation, we notice several \textbf{inefficiencies} in LightningSim, as shown in Fig.~\ref{fig:lightningsimv2}. \ul{Inefficiency 1}: LightningSim ignores optimization opportunities for common HLS design patterns during simulation, e.g., loops, resulting in repeated work. 
\ul{Inefficiency 2}: The stall calculation in LightningSim involves massive redundant computation and thus is inefficient.
It uses an event-based simulator, which repeatedly checks the ``events'' inside modules (more details in Sec~\ref{sec:LS-limitation}), resulting in slow simulation for large designs.
As presented in LightningSim, for medium-size designs such as FlowGNN~\cite{flowgnn}, it takes tens of seconds to simulate, while for large designs, it can take up to several tens of minutes.
\ul{Inefficiency 3}: Even though LightningSim applies incremental re-simulation for hardware parameter changes, it still needs to read and modify the raw list-based execution traces,
making the design space exploration (DSE) slow and inefficient for parallel computing using multicore. For instance, we observe that for a complicated HLS design, a first-order Implicit Neural Representation (INR) architecture with 238 FIFOs~\cite{inrarch}, each DSE iteration takes 4 minutes to visit only one design point, making it impossible to fully explore the optimal FIFO depths. 
A second-order INR architecture has more than 900 FIFOs and each design point takes more than 30 minutes to simulate using LightingSim, exposing great challenges for scalable DSE.

Addressing these limitations, we propose \textbf{LightningSimV2}. It achieves much faster simulation speed, scales better to larger designs, and enables much faster DSE, thanks to \textbf{three key innovations}, as depicted in Fig.~\ref{fig:lightningsimv2}.
We summarize our contributions and key features of LightningSimV2 as follows.

\begin{itemize}
    \item \textbf{Innovation 1: Optimized trace generation.} LightningSimV2 employs static analysis to identify repetitive design patterns, e.g., loop structures, and avoid tracing every iteration of these loops.
    The generated execution traces are much smaller and thus are faster to analyze.

    \item \textbf{Innovation 2: Optimized and \textit{decoupled} stall calculation.} Instead of event-based simulation in LightningSim, LightningSimV2 adopts a novel and much faster graph-based simulation technique.
    It decouples stall calculation into two steps: dynamic simulation graph construction and graph traversal.
    The first step constructs a simulation graph by scanning each simulation event \textit{only once}, avoiding \textit{repeatedly checking} the same events as LightningSim does.
    The second step applies a superfast graph traversal for the final cycle count calculation.
    Using the graph-based simulation, LightningSimV2 achieves up to 6.4\texttimes{} speed up for the trace analysis step.

    \item \textbf{Innovation 3: Optimized and \textit{parallelizable} DSE.} Our decoupled graph-based simulation technique also enables efficient DSE for HLS dataflow designs.
    A great feature of our proposed simulation graph is that it can capture unknown dependencies during graph construction; such dependencies rely on actual hardware parameters, e.g., FIFO depths. Thus the simulation graph is \textit{hardware-agnostic and does not need to change}.
    When performing DSE for FIFO depths, only the lightweight graph traversal step will be executed without modifying the simulation graph.
    This is not only efficient but more importantly, it enables high-throughput evaluation of multiple design points simultaneously by employing \textit{parallel computing} with little memory and communication overhead. 

    \item \textbf{Evaluation and open-source.} All of our proposed improvements are completely lossless. LightningSimV2 is 100\% accurate compared with LightningSim and thus is 99.9\% accurate to RTL simulation.
    It is up to 3.52\texttimes{} faster than LightningSim for end-to-end simulation, particularly for large designs. 
    In addition, LightningSimV2 shows superior DSE efficiency:
    its incremental stall calculation is up to 577\texttimes{}, 121\texttimes{} on average, faster than that in LightningSim.
    The code is open-source on GitHub.
    
\end{itemize}

\section{Background and Motivations}

\subsection{HLS Simulation}

Several prior works identify opportunities to speed up HLS simulation. FastSim~\cite{fastsim} uses the Verilog code synthesized by HLS to generate an equivalent C++ model to handle frequently used constructs, particularly the finite state machines (FSMs), for faster simulation. FLASH~\cite{flash} takes the approach of combining the input HLS source code with scheduling information from the synthesis process to generate a C++ model for computing latency estimates that cycle-accurately model FIFO communication. 
\textbf{LightningSim}~\cite{lightningsim} is another state-of-the-art HLS simulator, using a novel trace-based approach with up to two orders of speedup over RTL simulation.
We briefly introduce LightningSim approaches and their limitations.

\subsection{LightningSim Workflow}
\label{sec:LSv1-flow}

Fig.~\ref{fig:lightningsimv2} shows LightningSim's two decoupled stages, \textit{IR trace generation} and \textit{IR trace analysis}.
The trace analysis stage further has two steps, \textit{schedule resolution} and \textit{stall calculation}.

\begin{figure*}
    \centering
    \includegraphics[width=\linewidth]{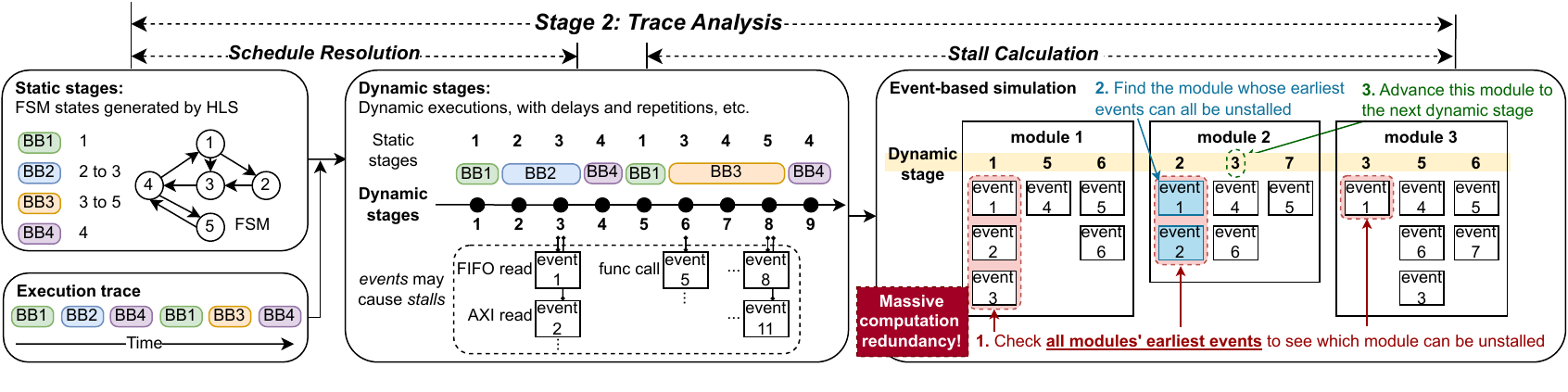}
    \caption{Illustration of the trace analysis stage in LightningSim, including schedule resolution and stall calculation. Schedule resolution correlates static stages to dynamic stages. Stall calculation uses an event-based simulation to calculate the final clock count, which involves massive computation redundancy since the same events may be checked repeatedly, and the number of events can be large.}
    \label{fig:stages}
\end{figure*}

\noindent\textbf{Trace generation.}
First, LightningSim compiles and instruments an HLS testbench to produce an execution trace of simulation events that determine hardware execution time. This trace includes a list of LLVM basic blocks (BBs) in their execution order that allow full reconstruction of exactly which LLVM instructions are executed and when. It also includes ancillary data about the simulated off-chip AXI memory access patterns and the read/write patterns of internal FIFO buffers.
The output of this step is a dumped file of the execution trace of BBs; an example is shown in Fig.~\ref{fig:stages} left bottom block.%

\noindent\textbf{Schedule resolution.}
Second, LightningSim analyzes this generated execution trace and correlates it with static scheduling data produced by the HLS tool to capture its dynamic behavior, as shown in Fig.~\ref{fig:stages}.
During this process, LightningSim creates a list of \textit{events} for each module.
An event indicates the interaction between the module and external signals, e.g., off-chip memory access, FIFO read/write, and function calls, which may cause modules to stall.

Each event has a \textit{static stage} after HLS synthesis and is mapped to one specific \textit{dynamic stage} after schedule resolution.
In this example, events 1 and 2 are mapped to dynamic stage 3, and event 5 is mapped to dynamic stage 6.
The dynamic stage represents the order in which the event occurs, accounting for repeated executions.
For instance, two events occurring in two different iterations of a loop would have the same static stage but different dynamic stages.
The output of this step is a set of event lists, each module with one list.

\noindent\textbf{Stall calculation.}
Once these events and their dynamic stages are determined, LightningSim uses a custom \textit{stall calculation} engine to iterate over the events, modeling actual hardware status as it goes, e.g., FIFO full/empty. The goal of stall calculation is to determine whether each event causes a stall and, if so, for how long, to precisely capture the design cycle count.
The output is the actual cycle count of the entire design.

The performance of stall calculation is particularly important:
it not only correctly calculates the actual design cycle count, but more importantly, its speed is crucial for efficient DSE of HLS dataflow designs.
If an HLS design has multiple FIFOs, optimally determining the depth of each FIFO is essential to hardware resources and performance but is challenging.
Using Vitis HLS, each time one FIFO depth is changed, a full-blown RTL simulation is needed to get its new cycle count.
In contrast, LightningSim can \textit{incrementally} re-execute the stall calculation step only; thus it is much more efficient than using RTL simulation.

\subsection{LightningSim Limitations}
\label{sec:LS-limitation}

Despite the promising features of LightningSim, we identify two limitations that may largely hurt its simulation speed and thus hinder efficient DSE for HLS designs.

\subsubsection{Fixed-Bound Loops}

HLS designs frequently feature \textit{fixed-bound} loops and pipelines with iteration counts known at compile-time. LightningSim processes these repetitive iterations one by one during trace generation, committing a large volume of redundant work and resulting in larger trace files. Imagine a loop with tripcount one million---LightningSim will generate one million repetitive BBs in its trace.

Prior work exists in other domains that uses static analysis techniques up-front to make dynamic analysis more efficient. For instance, \textsc{Cypress}~\cite{cypress} applies this concept to communication trace compression. It uses static analysis of a target program to construct a communication trace template, which means that dynamic analysis can trace only the information needed to ``fill in'' the template.

\subsubsection{Event-Based Simulation}
\label{sec:event-based-sim}

As discussed earlier, during schedule resolution, LightningSim processes the execution trace to generate lists of \textit{events}; then, during stall calculation, LightningSim adopts an event-based simulator to progressively check the behavior of events and modules.
\textit{This approach involves massive computation redundancy}.

The rightmost block in Fig.~\ref{fig:stages} shows an example of the simulation process in LightningSim.
In this example, there are three active modules, each contains a pool of events, and each event is mapped to a dynamic stage inside that module.
First, LightningSim will check all the events of all active modules to see whether they can be unstalled (the events annotated by red dashed circles). For instance, a FIFO read event will be unstalled if the FIFO is no longer empty.
Second, a module can only be unstalled if all its events are unstalled for its earliest dynamic stage (the events annotated by green).
Third, this module will be advanced to the next dynamic stage (module 2).
These three steps will be repeatedly executed till the simulation finishes.
During this process, the first step contains massive computation redundancy, since the same event may be visited multiple times. For example, after module 2 advances from dynamic stage 2 to 3, the same sets of events in modules 1 and 3 need to be checked again, since some events may change their stalling status due to event dependencies from module 2.
This can be wasteful, especially if there are many active modules but only one of them is unstalled per iteration. Most of the modules will not have changed state, but LightningSim must still re-check their events.
In addition, before simulating each module, LightningSim must sort all its events by their dynamic stages. This can cause high overhead for large designs since sorting complexity increases superlinearly with the event count.

The \textbf{fundamental limitation} of this approach is that events are maintained as lists, and the dependencies between events are unknown during static analysis and must be captured on-the-fly according to FIFO depths, off-chip memory accesses, and function calls. 
As an example, assume a FIFO depth is 3 and a consumer module is reading from it; to decide whether this read event can be unstalled, we must check the \textit{current} FIFO depth on-the-fly as well as the producer module's write event.
This requires repetitive scanning of modules and events.

This limitation raises a question: \textit{how can we dynamically capture such dependencies with only a one-time scan?}
In addition, since graphs are natural to capture dependencies, can we use a graph-based simulator instead of using event lists?

In the remainder of the paper, we introduce the three key innovations addressing the three inefficiencies in LightningSim. 
Sec.~\ref{sec:fix-bound-loop} discusses innovation 1, optimized trace generation with static analysis for fixed-bound loops.
Sec.~\ref{sec:decoupled-stall} discusses innovation 2, optimized stall calculation, by adopting a graph-based simulation rather than event-based, with two decoupled steps, graph construction and graph traversal.
Sec.~\ref{sec:DSE} discusses innovation 3 by showcasing efficient DSE of FIFO depths for a large design, benefiting from the decoupled stall calculation.

\section{Static Analysis for Fixed-Bound Loops}
\label{sec:fix-bound-loop}

\subsection{Challenges and Overview}
\noindent\textbf{Challenges.} Reducing the footprint of loops in the execution trace poses several challenges.

\begin{itemize}
    \item We must reliably identify fixed-bound loops and their corresponding tripcounts. We do not require users to annotate their code with accurate tripcounts.
    \item The static and dynamic start and end stage of such loops need to be correctly identified, for both pipelined and nonpipelined loops. 
    \item The order and stage information of events needs to be maintained for all events that occur within the loop---even for simple fixed-bound loops, this data may differ across loop iterations. 
\end{itemize}

\noindent\textbf{Overview.}
During the trace generation step, loops with fixed bounds are identified, the tripcount is found, and the trace logging of the basic blocks in the loop is removed. Trace logging for the fixed loop is then inserted. This greatly reduces the number of trace file lines generated for each fixed-bound loop by only logging each basic block in the loop once. %

During the trace analysis step, when a loop is encountered, the loop tripcount and basic blocks are parsed and the loop dynamic start and end stages are calculated. Then, all AXI and FIFO events %
are handled for that loop. Thus we only process each basic block in the loop once instead of every iteration. 

\subsection{Step 1: Trace Generation}
\noindent\textbf{Simple fixed-bound loop identification.} A simple fixed-bound loop is identified by a loop containing two basic blocks (a header and a latch) that has a constant tripcount, as identified by %
LLVM loop analysis passes~\cite{llvm}. This is reliable and does not need an accurate user-defined tripcount. 

\noindent\textbf{Trace generation modification.} The trace generation is then modified to log each basic block in the loop only once by moving the log statements to the pre-header of the loop and removing the log statements from each basic block in the loop. Events will then be logged as normal, and an end loop statement is added to the exit block of the loop.

\subsection{Step 2: Trace Analysis}
\noindent\textbf{Loop end dynamic stage.} At the trace analysis step, the loop basic blocks are parsed, setting the dynamic start and end stage of the loop. In a pipelined region, the dynamic end stage of the loop is set to:
\(\text{dynamic start stage} + \text{loop overlap length} + \text{II} \times (\text{tripcount} -1) \). In a non-pipelined region, the end stage of the loop is set to: \(\text{dynamic start stage} + \text{loop overlap length} + (\text{loop overlap length} + 1) \times (\text{tripcount} - 1)\). However, the loop header basic block always executes one extra time, so the dynamic end stage is adjusted to reflect the extra execution. 

\noindent\textbf{Events.} To maintain correctness, we still log all events in the execution trace. As events in the loop are analyzed, the LightningSimV2 schedule resolver loops through the corresponding blocks, and the iteration number of the loop is used to assign the dynamic start and end stages of each event instance.

\begin{figure*}
    \centering
    \includegraphics[width=\linewidth]{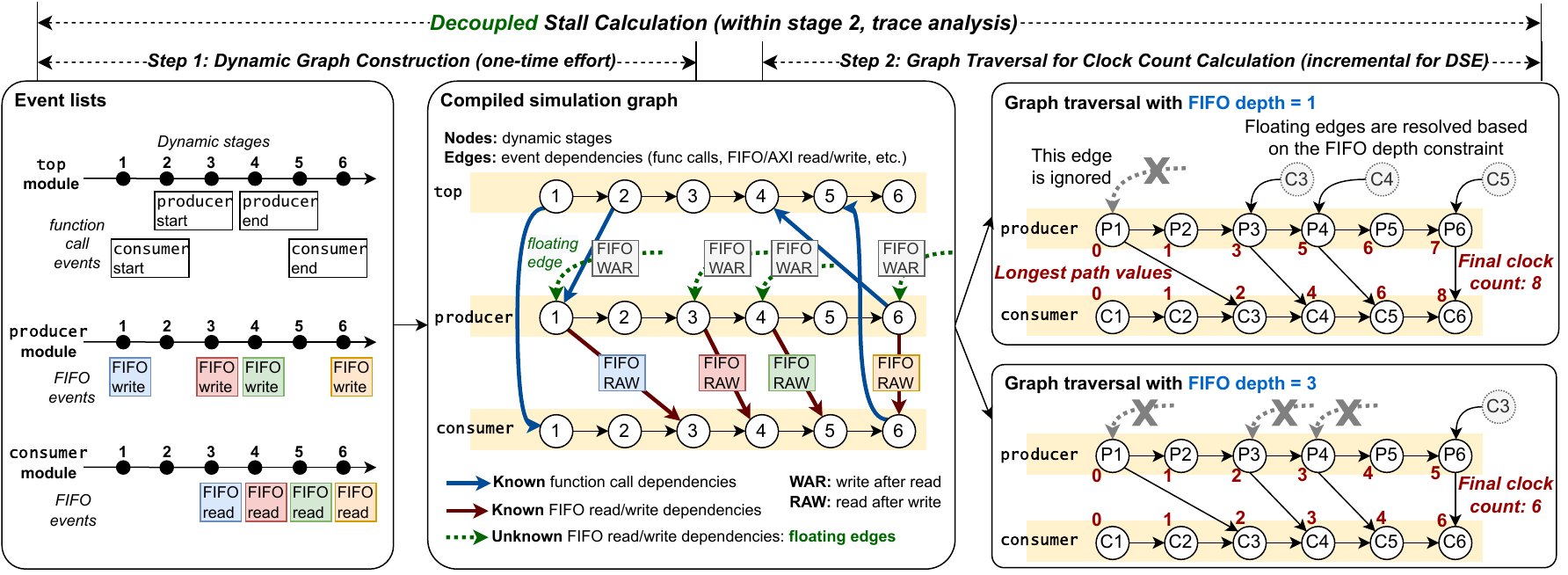}
    \caption{Illustration of the decoupled stall calculation in LightningSimV2, including two steps. Step 1: dynamic graph construction, which is a one-time effort. Step 2: graph traversal for clock count calculation, which is extremely lightweight and can be applied for DSE, e.g., with different FIFO depths. }
    \label{fig:graph-example}
\end{figure*}

\section{Decoupled Stall Calculation}
\label{sec:decoupled-stall}

\subsection{Graph-based Simulation: Challenges and Overview}

As discussed in Sec.~\ref{sec:event-based-sim}, the limitation of event-based simulation is the lack of known dependencies between events, thus requiring repeated checks.
This naturally motivates the usage of graphs to capture and record those dependencies.

\noindent\textbf{Challenges.}
Using graphs is a well-established approach for many applications, such as task scheduling~\cite[p.~469]{skiena2008algorithm} and deadlock detection~\cite{hilbrich2009graph}. INR-Arch~\cite{inrarch} first proposed the use of graphs to detect deadlocks and calculate latency estimates in a FIFO-heavy HLS design. However, their graph construction approach is specific to their HLS design and does not generalize, especially to those with off-chip memory accesses.

Aiming to propose a general-purpose and scalable simulator based on a \textit{simulation graph}, we encounter specific challenges.

\begin{itemize}

    \item \textbf{C1:} The dependencies are partially determined by hardware parameters, such as FIFO depths, which can only be modeled by checking the hardware resources at each time step as the simulation progresses. \textit{How can we model unknown dependencies during the graph construction?}

    \item \textbf{C2:} Event dependencies differ according to different hardware parameters, resulting in a different dependency graph, so
    changing the hardware parameters may require constructing a new graph each time.
    This can be prohibitively expensive if a DSE of hardware parameters is needed, e.g., deciding the optimal FIFO depths within a large design.
    \textit{How can we avoid constructing a new graph from scratch for each set of new design parameters?}\label{sec:challenge-fifo-depths-reconstruction}

    \item \textbf{C3:} To reduce the size of the simulation graph and make it easy to analyze, the graph should use compressed sparse row (CSR) format. CSR is most performant when frequently querying a node's predecessors, which best suits the goal of simulation. However, it requires that the predecessors of a node are known during construction time. Given the dynamic nature of the potentially unknown dependencies, \textit{how can we construct the graph in CSR without knowing all the predecessors of the nodes?}

    \item \textbf{C4:} If the HLS design and/or the number of events is large, the graph will contain a huge number of nodes, which can be slow to construct and analyze. \textit{How can we reduce the graph size so it is memory efficient?}

\end{itemize}

\noindent\textbf{Overview.}
Fig.~\ref{fig:graph-example} illustrates the overall flow of our innovative graph-based stall calculation.
It has two decoupled steps: simulation graph construction and graph traversal.
Graph construction is a one-time effort that scans the event lists \textit{only once} while keeping the unresolved dependencies, which we call \textit{floating edges}.
Graph traversal is a lightweight process to compute the cycle count on the simulation graph while resolving the dependencies.
Given different hardware parameters, the simulation graph remains the same but only the graph traversal is different. 
In this example, we demonstrate two cases, a FIFO depth being 1 and 3: both cases use the same simulation graph but have different traversal results and floating edge resolution.

Using our proposed algorithm, the decoupled approach addresses \textbf{C2}; the floating edge approach addresses \textbf{C1}.
We further propose a \textit{node pending/commit} technique to address \textbf{C3} and node/edge elimination to address \textbf{C4}.

\subsection{Step 1: Dynamic Graph Construction}

\begin{figure}
    \centering
    \includegraphics[width=\linewidth]{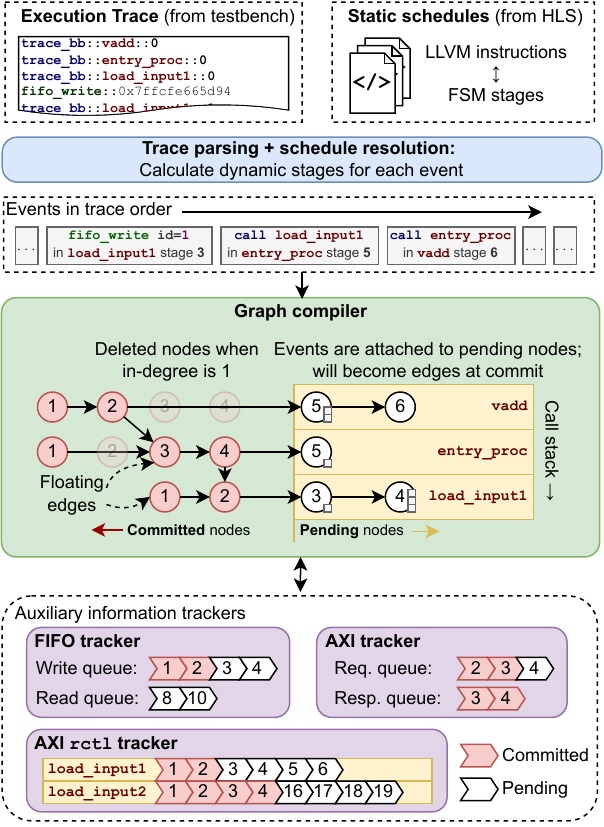}
    \caption{The LightningSimV2 graph compiler architecture. Trace resolution produces a stream of events, each of which is timestamped with a dynamic stage in its corresponding module. These events are attached to pending nodes within the call stack and added to tracking structures. Eventually, when a pending node is committed, its events are used to update tracking structures that also determine what edges need to be created in the graph.}
    \label{fig:graph-compiler}
\end{figure}

The simulation graph is built based on dynamic stages and events.
The concept of dynamic stages is introduced in Sec.~\ref{sec:LSv1-flow} and illustrated in Fig.~\ref{fig:stages}.
We will introduce the definition of an event, the simulation graph structure, the process of building the graph, and data structures to store auxiliary information to assist graph construction.

\subsubsection{Event}
Events are the fundamental building block of simulations in both LightningSim and LightningSimV2. An event represents the time step at which one module may affect another.
Each event is associated with a time at which it occurs, expressed as a dynamic stage of the module in which it exists. Multiple events may co-exist at a dynamic stage.

In LightningSimV2, the input to the graph construction process is a stream of events that are generated from the execution trace, as indicated in the top half of Fig.~\ref{fig:graph-compiler}. LightningSimV2 defines seven different types of events:
\begin{itemize}
    \item The \ul{call} event occurs at the start of a function call in the execution trace. Events following the call %
    will compose a subgraph within the simulation graph. In hardware, this equates to a submodule's start signal being asserted.
    
    \item The \ul{return} event indicates that the current function call has reached completion. This represents the hardware module asserting its done signal to unblock its parent module. All pending nodes in the current module will be \textit{committed} (will be discussed later in this section) and we return to building the subgraph for the parent module.
    
    \item \ul{FIFO read} and \ul{FIFO write} events correspond to a read or write of a FIFO stream within the current function. In hardware, a module trying to read an empty FIFO buffer is stalled until the buffer is not empty; a module trying to write a full FIFO buffer is stalled until it is not full.
    
    \item \ul{AXI read request} and \ul{AXI read} events are generated in the execution trace during a load from off-chip memory over an AXI interface. An \ul{AXI read request} must precede one or more \ul{AXI reads}, 
    and a minimum delay in terms of cycles is enforced in our simulator between an \ul{AXI read request} and its first \ul{AXI read}, mimicking C/RTL co-sim.

    \item Similarly, \ul{AXI write request}, \ul{AXI write}, and \ul{AXI write response} events are generated during stores to off-chip memory. A minimum number of cycles between \ul{AXI writes} and \ul{AXI write responses} is also enforced.
    
\end{itemize}

\subsubsection{Simulation Graph Structure}
Our proposed simulation graph is composed of {nodes} that are either committed or pending, {edges} with known source and target nodes, and {floating edges} with an unknown source.

\begin{itemize}
    \item The \ul{nodes} in the simulation graph 
represent dynamic stages within hardware modules, as shown in Fig.~\ref{fig:graph-example}.
Each node can be associated with multiple events.

\item The \ul{edges} represent known dependencies between events that are associated with the nodes---constraints that one event must ``happen before'' another event. Each edge has its meta-data to record the dependency type. \label{sec:what-are-nodes-and-edges} 

\item The \ul{floating edges} represent unknown FIFO write-after-read (WAR) dependencies, due to unknown FIFO depths. For a FIFO write event, it will be stalled if the FIFO is full, and thus a FIFO WAR dependency must be created for it.
However, we must avoid hardcoding such constraints into the graph; otherwise the graph must be rebuilt for any FIFO depth change.
Therefore, we create a FIFO WAR floating edge to each FIFO write event (except the first, which can never depend on a FIFO read) but leave its source unspecified. 
These floating edges will be resolved, i.e., the source will be specified, in the second step, graph traversal, according to specific FIFO depth values.
The proposed floating edge addresses the challenge of unknown dependencies (\textbf{C1}).

\end{itemize}

\noindent\ding{114} \textbf{Example.}
We first use a simple example in Fig.~\ref{fig:graph-example} to explain the graph structure before discussing detailed edge types. It depicts three modules: a \texttt{top} module that invokes a \texttt{producer} module writing to a FIFO stream and a \texttt{consumer} module reading from it. In each module, we first create a node for each dynamic stage, from 1 to 6. We notice a dependency between FIFO reads and writes: if a write does not occur before its corresponding read, the FIFO will be empty, and the read will stall until the write occurs. Thus, we create edges from FIFO writes to FIFO reads to enforce this dependency that ``writes must happen before reads''.
As shown in the compiled simulation graph, there are four such edges annotated as FIFO RAW (read after write).
Next, for each FIFO write event for the \texttt{producer} module, there must be a FIFO WAR dependency. Since the FIFO depth is unknown during graph construction, these dependency edges are floating.

Formally, the simulation graph contains six types of edges:
\begin{itemize}
    \item \ul{Control flow} edges are the most ubiquitous in the graph. They are used for two purposes. First, they enforce the implicit ordering constraints between consecutive dynamic stages of the same module. For example, since dynamic stage 2 of any module must always come after its dynamic stage 1, an edge must be created between these two nodes. Second, control flow edges are used at the start and end of each subcall, as depicted by the \texttt{top} module in Fig.~\ref{fig:graph-example}: since \texttt{producer} starts at \texttt{top}'s stage 2 and ends at its stage 4, blue arrows connect \texttt{top}'s stage 2 to the first stage of \texttt{producer} and the last stage of \texttt{producer} back to \texttt{top}'s stage 4.
    \item \ul{FIFO read-after-write (RAW)} edges connect each FIFO write to the same-numbered FIFO read. As previously mentioned, at the time that the \(n\)\textsuperscript{th} FIFO read would normally occur, the \(n\)\textsuperscript{th} FIFO write must have already occurred; otherwise the FIFO will be empty and the read will stall until the write occurs. Either situation can be modeled through a dependency from the write to the read.
    \item \ul{FIFO write-after-read (WAR)} edges model dependencies caused by a FIFO becoming full. If a FIFO has depth \(d\), after \(d\) writes to the FIFO, at least one read must occur before the \((d + 1)\)\textsuperscript{st} %
    write can occur. Therefore, for a FIFO of depth \(d\), we must create an edge to model that the \(n\)\textsuperscript{th} write depends on the \((n - d)\)\textsuperscript{th} read. 
    As discussed earlier, these edges are floating edges.
    \item \ul{AXI read} edges are used to create a dependency between a read request on an AXI interface and its first read transfer. Each AXI interface has a user-specified latency, e.g., 64 cycles; during simulation, the first read transfer is modeled as occurring at least this long (plus some fixed overhead) after the read request. \ul{AXI write response} edges model a similar dependency between the last write transfer on an AXI interface and the write response.
    \item \ul{AXI \texttt{rctl}} edges model a more complicated dependency that stems from the hardware implementation of AXI interfaces. Each AXI interface has its own internal FIFO buffer named \texttt{rctl} that buffers read requests; however, once this buffer becomes full, any subsequent read requests will be delayed until the active read request is fully processed and can be removed from the \texttt{rctl} buffer. The observed effect of this delay is that once the \texttt{rctl} buffer is full, the first read transfer of subsequent read requests stalls for longer than it normally would. Since a read request is removed from the \texttt{rctl} buffer after its last read transfer, a dependency is created by the last read transfer of a read request and the first read transfer of a request some time later.
\end{itemize}

\subsubsection{Graph Compilation Process} 
As discussed earlier, we prefer the simulation graph to be in CSR format, which is memory efficient and suitable for graph traversal.
However, we must address two unique challenges: \textbf{C3:} a node cannot be added to the graph until all its predecessors are known, and \textbf{C4:} the graph can be prohibitively large.

To address \textbf{C3}, we propose to keep nodes in two statuses: \textit{pending} (depicted in white in Fig.~\ref{fig:graph-compiler}) and \textit{committed} (depicted in red). Nodes are pending when their in-degree may change as more events arrive. Once a node's in-degree is known, it is committed, meaning that it is added to the CSR data structures.

As events stream from the execution trace, the graph compiler maintains a call stack of modules and a queue of pending nodes for each, corresponding to the latest dynamic stages in each module. For each event streamed in, the node corresponding to its dynamic stage is annotated with the event. Events sharing a dynamic stage are grouped onto one node.

The graph compiler \textit{commits} a node when it is certain that no more events in the execution trace will reference it. This is possible because of the nature of the schedule resolution process: after an event with static stage \(s\) and dynamic stage \(d\), no event after it in the execution trace can occur at a dynamic stage less than \((d - s)\). Therefore, we can always commit pending nodes whose dynamic stages are earlier than that, ensuring that the pending node queue for each module is never larger than the latest static stage encountered in that module.\label{sec:node-commit}

To address \textbf{C4}, we apply node elimination on the fly.
Specifically, for nodes with only one incoming edge, we entirely remove them from the graph by combining the single in-edge with each of its out-edges.\label{sec:node-elimination}

The proposed node pending/commit and elimination techniques can largely reduce graph size, as Sec.~\ref{sec:experiment} will show.

\subsubsection{Auxiliary Information Trackers during Graph Construction}
During graph construction, we need to keep track of FIFO status and AXI events and maintain their correct orders. We design the following three trackers, as depicted in Fig.~\ref{fig:graph-compiler}.

\begin{itemize}
    \item \ul{FIFO tracker.} For each FIFO in the design, we keep a FIFO tracker to observe any of its read or write events. This tracker serves two purposes. First, it provides each FIFO read or write event with a pairing key, queueing these keys so that each read is connected to a write using this pairing key. Second, the tracker also collects the permanent node indices of each read and write as they are committed; this is crucial for resolving floating WAR edges to a FIFO read during graph traversal.

    \item \ul{AXI tracker.} 
 AXI trackers similarly track all events for an AXI interface, providing events with the keys corresponding to AXI read and AXI write response edges. Unlike the FIFO tracker, no queue is necessary because the source and destination of each of these edge types always appear back-to-back in the execution trace with no other AXI events separating them. Instead, an AXI read request simply creates the edge key to be used by the following read, and an AXI write (the last in a write transaction) creates the edge key to be used by the following write response.

    \item  \ul{AXI \texttt{rctl} tracker.} 
Modeling the dependencies caused by the AXI \texttt{rctl} buffer is significantly more complex than for FIFO and other AXI dependencies. In particular, dependencies caused by the AXI \texttt{rctl} buffer rely on a global ordering of all AXI read requests across the entire design, which must be specially handled.
Furthermore, if a read request occurs before a subcall, any AXI transfers occurring within the subcall must be correctly ordered with respect to the read request in the parent module.

The AXI \texttt{rctl} tracker maintains the correct global ordering using two key principles. First, pending nodes are guaranteed to be committed in the order of their dynamic stages, so the \texttt{rctl} tracker uses commit order, rather than execution trace order, to put read requests into the \texttt{rctl} buffer. Second, to handle inter-module dependencies, the \texttt{rctl} tracker maintains \texttt{rctl} buffer models separately for each module in the design and splices them into the correct order when the module \ul{return} event is encountered.

\end{itemize}

\subsection{Step 2: Graph Traversal}

Once the simulation graph has been constructed, traversing the graph to calculate the simulation cycle count becomes a very simple and fast process of finding the longest path through the graph. This can be done by iterating over every node in topologically sorted order and assigning a cycle count to the node, which is the maximum value of its predecessor cycle count plus the corresponding incoming edge delay.
The value of the end node in the graph is the final cycle count of the HLS design.

During graph traversal, the floating edges will be resolved, i.e., the source node will be determined, according to the FIFO depth parameters. 
The only type of floating edge used in the LightningSimV2 is the FIFO WAR edge, pointing to FIFO write events.
Assume the FIFO between a producer and a consumer has a depth of $d$.
If the floating edge destination is the \(n\)\textsuperscript{th} FIFO write event, then the source must be resolved to \((n-d)\)\textsuperscript{th} FIFO read, indicating that the FIFO is full and cannot be written unless a FIFO read is executed. If \(n\) is smaller than \(d\), the FIFO write will have no dependency on any FIFO read, and thus the floating edge can be ignored entirely.

Notably, this graph traversal \textit{does not} change the simulation graph structure, making it ideal for incremental DSE and parallel computing.

\noindent\ding{114} \textbf{Example.}
We showcase two graph traversals in Fig.~\ref{fig:graph-example}, with FIFO depths being 1 and 3.
When the FIFO depth is 1, the last three FIFO WAR edges are resolved to their respective FIFO reads. Specifically, P3's incoming floating edge is resolved to C3, because P3 is the second FIFO write and C3 is the first FIFO read, and with FIFO depth being 1, P3 will be stalled by C3 since the FIFO is full now.
The first FIFO write is never stalled since the FIFO depth is always larger than zero.
When the FIFO depth is 3, only the last floating edge for P6 is resolved to C3 while the first three are ignored.
With the floating edges resolved, the longest path values, annotated as red, are computed accordingly.

Adapting an established linear-time algorithm for topological sorting~\cite[p.~573]{cormen2022introduction}, we implement our simulation by performing a depth-first postorder traversal on the reverse of the simulation graph. We start at the end node of the simulation and perform a depth-first search along the reverse direction of each directed edge in the graph, visiting each node's predecessors rather than its successors. After visiting all of a node's predecessors, we update that node's clock cycle according to the maximum of each predecessor node's clock cycle combined with the delay along that predecessor edge.

We highlight that if this depth-first traversal encounters a cycle in the graph, it indicates that the design will deadlock---two (or more) nodes must happen before each other, which means they will both stall indefinitely in hardware. Thus our simulator can perform deadlock detection with no added cost. %

\section{Experimental Results}
\label{sec:experiment}

We evaluate the effectiveness of our proposed techniques through several comprehensive experiments. All experiments were evaluated on a 64-core machine using an Intel Xeon Gold 6226R x86-64 CPU and 502 GiB of RAM, running Red Hat Enterprise Linux Server 7.9 and tested using AMD Vitis HLS 2021.1. We reproduce the results of LightningSim on our system using the open-source code.\footnote{\url{https://github.com/sharc-lab/LightningSim/tree/v0.1.0}}

\subsection{Comparisons Against LightningSim}\label{sec:original-comparisons}

\begin{table*}
    \caption{Comparisons of LightningSimV2 over LightningSim.}
    \label{tab:results}
    \centering
    \setlength{\tabcolsep}{1.805pt}
    \renewcommand*{\arraystretch}{0.925}
    \footnotesize
    \begin{tabular}{l|cc|cccc|cc|cccc|cc|cccc}
        \toprule
        &&& \multicolumn{6}{c|}{\textbf{LightningSim}} & \multicolumn{6}{c|}{\textbf{LightningSimV2}} \\
        &&& \multicolumn{4}{c|}{\textbf{Time}} & \multicolumn{2}{c|}{\textbf{Trace}} & \multicolumn{4}{c|}{\textbf{Time}} & \multicolumn{2}{c|}{\textbf{Trace}} \\
        & \multicolumn{2}{c|}{\textbf{Cycles}} & \textbf{Total} & \textbf{TG} & \textbf{TA} & \textbf{Incr.} & \textbf{Line} & \textbf{Size} & \textbf{Total} & \textbf{TG} & \textbf{TA} & \textbf{Incr.} & \textbf{Line} & \textbf{Size} & \multicolumn{4}{c}{\textbf{Speedup}} \\
        \textbf{Benchmark} & \textbf{Cosim} & \textbf{LS/LSv2} & (s) & (s) & (ms) & (ms) & \textbf{Count} & (bytes) & (s) & (s) & (ms) & (ms) & \textbf{Count} & (bytes) & \textbf{Overall} & \textbf{TG} & \textbf{TA} & \textbf{Incr.} \\
        \midrule
        Fixed-point square root~\cite{xilinx2021basic} & 30 & 30 & 5.00 & 5.00 & 3.91 & 0.12 & 6.3K & 151K & 4.97 & 4.97 & 2.80 & 0.00 & 700 & 18K & 1.01\texttimes{} & 1.01\texttimes{} & 1.39\texttimes{} & 77.3\texttimes{} \\
        FIR filter~\cite{xilinx2021basic} & 172 & 172 & 2.38 & 2.37 & 5.32 & 0.12 & 128K & 2.4M & 2.43 & 2.43 & 6.02 & 0.00 & 128K & 2.4M & 0.98\texttimes{} & 0.98\texttimes{} & 0.88\texttimes{} & 71.4\texttimes{} \\
        Fixed-point window conv~\cite{xilinx2021basic} & 35 & 35 & 3.56 & 3.56 & 2.15 & 0.08 & 67 & 1.7K & 3.69 & 3.69 & 2.31 & 0.00 & 7 & 182 & 0.97\texttimes{} & 0.97\texttimes{} & 0.93\texttimes{} & 53.3\texttimes{} \\
        Floating point conv~\cite{xilinx2021basic} & 35 & 35 & 2.37 & 2.35 & 17.2 & 0.12 & 67 & 1.7K & 2.42 & 2.42 & 1.92 & 0.00 & 7 & 182 & 0.98\texttimes{} & 0.97\texttimes{} & 8.96\texttimes{} & 74.7\texttimes{} \\
        Arbitrary precision ALU~\cite{xilinx2021basic} & 36 & 36 & 2.09 & 2.09 & 3.24 & 0.07 & 9 & 252 & 2.12 & 2.12 & 2.38 & 0.00 & 9 & 252 & 0.99\texttimes{} & 0.99\texttimes{} & 1.36\texttimes{} & 47.9\texttimes{} \\
        Parallel loops~\cite{xilinx2021basic} & 32 & 32 & 2.32 & 2.31 & 5.55 & 0.14 & 131 & 5.2K & 2.34 & 2.34 & 4.68 & 0.00 & 131 & 5.2K & 0.99\texttimes{} & 0.99\texttimes{} & 1.19\texttimes{} & 65.4\texttimes{} \\
        Imperfect loops~\cite{xilinx2021basic} & 34 & 34 & 2.38 & 2.37 & 8.25 & 0.12 & 44 & 1.8K & 2.24 & 2.23 & 8.73 & 0.00 & 8 & 276 & 1.06\texttimes{} & 1.06\texttimes{} & 0.95\texttimes{} & 69.6\texttimes{} \\
        Loop with max bound~\cite{xilinx2021basic} & 31 & 31 & 2.24 & 2.24 & 2.63 & 0.11 & 2.1K & 56K & 2.25 & 2.25 & 1.94 & 0.00 & 224 & 6.1K & 0.99\texttimes{} & 0.99\texttimes{} & 1.36\texttimes{} & 68.8\texttimes{} \\
        Perfect nested loops~\cite{xilinx2021basic} & 406 & 406 & 2.26 & 2.25 & 7.39 & 0.21 & 1.3K & 30K & 2.27 & 2.26 & 5.85 & 0.00 & 1.3K & 30K & 0.99\texttimes{} & 0.99\texttimes{} & 1.26\texttimes{} & 134\texttimes{} \\
        Pipelined nested loops~\cite{xilinx2021basic} & 405 & 405 & 2.25 & 2.24 & 4.86 & 0.16 & 16K & 402K & 2.23 & 2.23 & 1.56 & 0.00 & 140 & 3.7K & 1.01\texttimes{} & 1.01\texttimes{} & 3.12\texttimes{} & 95.8\texttimes{} \\
        Sequential accumulators~\cite{xilinx2021basic} & 32 & 32 & 2.29 & 2.28 & 5.72 & 0.16 & 131 & 5.5K & 2.29 & 2.28 & 3.88 & 0.00 & 131 & 5.5K & 1.00\texttimes{} & 1.00\texttimes{} & 1.48\texttimes{} & 76.9\texttimes{} \\
        Accumulators + asserts~\cite{xilinx2021basic} & 33 & 33 & 2.31 & 2.31 & 5.61 & 0.15 & 103 & 5.0K & 2.30 & 2.29 & 4.41 & 0.00 & 103 & 5.0K & 1.01\texttimes{} & 1.01\texttimes{} & 1.27\texttimes{} & 75.3\texttimes{} \\
        Accumulators + dataflow~\cite{xilinx2021basic} & 31 & 31 & 2.31 & 2.30 & 6.23 & 0.19 & 131 & 3.5K & 2.29 & 2.29 & 3.96 & 0.00 & 131 & 3.5K & 1.01\texttimes{} & 1.00\texttimes{} & 1.58\texttimes{} & 93.7\texttimes{} \\
        Static memory example~\cite{xilinx2021basic} & 66 & 66 & 2.24 & 2.24 & 4.37 & 0.17 & 4.2K & 192K & 2.18 & 2.18 & 3.47 & 0.00 & 480 & 17K & 1.03\texttimes{} & 1.03\texttimes{} & 1.26\texttimes{} & 78.8\texttimes{} \\
        Pointer casting example~\cite{xilinx2021basic} & 408 & 408 & 2.18 & 2.17 & 5.18 & 0.14 & 819 & 25K & 2.15 & 2.15 & 1.85 & 0.00 & 7 & 209 & 1.01\texttimes{} & 1.01\texttimes{} & 2.80\texttimes{} & 92.4\texttimes{} \\
        Double pointer example~\cite{xilinx2021basic} & 25 & 25 & 2.05 & 2.04 & 5.05 & 0.16 & 490 & 21K & 2.14 & 2.14 & 5.18 & 0.00 & 170 & 6.0K & 0.96\texttimes{} & 0.96\texttimes{} & 0.98\texttimes{} & 79.3\texttimes{} \\
        AXI4 master~\cite{xilinx2021basic} & 178 & 177 & 2.16 & 2.16 & 7.93 & 1.53 & 416 & 14K & 2.19 & 2.18 & 4.10 & 0.00 & 128 & 3.3K & 0.99\texttimes{} & 0.99\texttimes{} & 1.93\texttimes{} & 568\texttimes{} \\
        AXIS w/o side channel~\cite{xilinx2021basic} & 52 & 51 & 2.04 & 2.04 & 2.10 & 0.09 & 103 & 2.0K & 2.06 & 2.06 & 1.21 & 0.00 & 7 & 158 & 0.99\texttimes{} & 0.99\texttimes{} & 1.73\texttimes{} & 61.5\texttimes{} \\
        Multiple array access~\cite{xilinx2021basic} & 252 & 252 & 2.24 & 2.24 & 2.79 & 0.14 & 255 & 8.2K & 2.18 & 2.18 & 1.64 & 0.00 & 7 & 213 & 1.03\texttimes{} & 1.03\texttimes{} & 1.70\texttimes{} & 92.4\texttimes{} \\
        Resolved array access~\cite{xilinx2021basic} & 131 & 131 & 2.24 & 2.23 & 6.39 & 0.20 & 256 & 14K & 2.20 & 2.19 & 3.68 & 0.00 & 8 & 332 & 1.02\texttimes{} & 1.02\texttimes{} & 1.74\texttimes{} & 112\texttimes{} \\
        URAM with ECC~\cite{xilinx2021basic} & 115 & 115 & 2.29 & 2.29 & 6.28 & 0.26 & 70 & 3.2K & 2.21 & 2.21 & 4.86 & 0.00 & 22 & 828 & 1.04\texttimes{} & 1.04\texttimes{} & 1.29\texttimes{} & 114\texttimes{} \\
        Fixed-point Hamming~\cite{xilinx2021basic} & 259 & 259 & 2.42 & 2.41 & 5.76 & 0.13 & 515 & 13K & 2.37 & 2.36 & 1.46 & 0.00 & 7 & 189 & 1.02\texttimes{} & 1.02\texttimes{} & 3.94\texttimes{} & 87.8\texttimes{} \\
        Unoptimized FFT~\cite{kastner2018parallel} & 261,781 & 261,150 & 2.91 & 2.65 & 260 & 78.0 & 39K & 1.2M & 2.78 & 2.68 & 107 & 0.82 & 39K & 1.2M & 1.05\texttimes{} & 0.99\texttimes{} & 2.43\texttimes{} & 94.8\texttimes{} \\
        Multi-stage FFT~\cite{kastner2018parallel} & 3,770 & 3,772 & 2.66 & 2.62 & 42.4 & 0.83 & 5.7K & 149K & 2.67 & 2.60 & 66.7 & 0.00 & 5.7K & 149K & 1.00\texttimes{} & 1.01\texttimes{} & 0.64\texttimes{} & 210\texttimes{} \\
        Huffman encoding~\cite{kastner2018parallel} & 10,283 & 10,272 & 2.83 & 2.59 & 233 & 3.94 & 25K & 987K & 2.63 & 2.53 & 102 & 0.01 & 24K & 935K & 1.07\texttimes{} & 1.02\texttimes{} & 2.29\texttimes{} & 266\texttimes{} \\
        Matrix multiplication~\cite{kastner2018parallel} & 1,036 & 1,036 & 2.64 & 2.63 & 11.6 & 0.24 & 2.1K & 43K & 2.61 & 2.61 & 3.51 & 0.00 & 7 & 172 & 1.01\texttimes{} & 1.01\texttimes{} & 3.31\texttimes{} & 155\texttimes{} \\
        Parallelized merge sort~\cite{kastner2018parallel} & 131 & 131 & 2.23 & 2.22 & 14.1 & 0.34 & 994 & 26K & 2.27 & 2.26 & 9.62 & 0.00 & 994 & 26K & 0.98\texttimes{} & 0.98\texttimes{} & 1.47\texttimes{} & 90.4\texttimes{} \\
        Vector add with stream~\cite{xilinx2022vitis} & 4,261 & 4,261 & 4.95 & 4.29 & 658 & 407 & 70K & 2.3M & 4.48 & 4.19 & 286 & 1.00 & 70K & 2.2M & 1.10\texttimes{} & 1.02\texttimes{} & 2.31\texttimes{} & 406\texttimes{} \\
        FlowGNN GIN~\cite{flowgnn} & 260,359 & 260,337 & 48.3 & 25.6 & 23\,s & 2.2\,s & 1.9M & 169M & 28.9 & 20.0 & 8.8\,s & 3.80 & 1.9M & 169M & 1.67\texttimes{} & 1.28\texttimes{} & 2.57\texttimes{} & 577\texttimes{} \\
        FlowGNN GCN~\cite{flowgnn} & 112,836 & 112,561 & 41.5 & 30.1 & 11\,s & 2.1\,s & 809K & 61M & 30.9 & 27.5 & 3.5\,s & 6.20 & 809K & 61M & 1.34\texttimes{} & 1.09\texttimes{} & 3.31\texttimes{} & 335\texttimes{} \\
        FlowGNN GAT~\cite{flowgnn} &  17,282 & 17,282 & 46.3 & 37.7 & 8.6\,s & 751 & 346K & 30M & 41.6 & 35.8 & 5.9\,s & 4.75 & 346K & 30M & 1.11\texttimes{} & 1.05\texttimes{} & 1.46\texttimes{} & 158\texttimes{} \\
        FlowGNN PNA~\cite{flowgnn} & 344,206 & 344,206 & 45.0 & 24.3 & 21\,s & 1.4\,s & 1.9M & 161M & 30.5 & 22.4 & 8.1\,s & 4.19 & 1.9M & 161M & 1.48\texttimes{} & 1.09\texttimes{} & 2.55\texttimes{} & 333\texttimes{} \\
        FlowGNN DGN~\cite{flowgnn} & 110,710 & 110,710 & 36.0 & 24.5 & 11\,s & 1.3\,s & 1.1M & 42M & 26.9 & 20.8 & 6.0\,s & 2.76 & 1.1M & 42M & 1.34\texttimes{} & 1.18\texttimes{} & 1.90\texttimes{} & 473\texttimes{} \\
        \bottomrule
    \end{tabular}

    \vspace{0.5em}

    \begin{tabular}{l}
        \footnotesize
        \textbf{Cosim Cycles:} Cycle count reported by C/RTL co-simulation.\quad
        \textbf{LS/LSv2 Cycles:} Cycle count reported by both LightningSim and LightningSimV2. \\
        \textbf{TG:} Time taken for trace generation.\quad
        \textbf{TA:} Time taken for trace analysis.\quad
        \textbf{Incr.:} Time taken for incremental simulation after changing FIFO depths. \\
        In all columns, K represents 1,000; M represents 1,000,000. \\
    \end{tabular}
\end{table*}

Table~\ref{tab:results} shows direct comparisons with the original LightningSim using the benchmark suite provided by the original LightningSim authors~\cite{lightningsim}. %
We compare the following metrics to evaluate the innovations of LightningSimV2.
\begin{itemize}
    \item Simulated cycle counts for simulation accuracy. These show that LightningSimV2 achieves 100\% accuracy with respect to the original LightningSim and 99.9\% accuracy with respect to C/RTL co-sim.

    \item \textbf{Total} end-to-end simulation time. On tiny designs, where trace generation dominates the total execution time and trace analysis takes under 0.1 seconds, LightningSimV2's performance is comparable to LightningSim.
However, when trace analysis takes a significant portion of the simulation, LightningSimV2 achieves considerable speedup of up to \textbf{1.67\texttimes{}} over the original LightningSim.

    \item Trace generation time (\textbf{TG}) and trace file size for fixed-bound loop optimization. 
The fixed-bound loop optimization technique contributes up to 1.28\texttimes{} speed up for trace generation.
More importantly, it reduces the amount of trace data significantly for multiple designs that are loop-heavy. For instance, LightningSim's matrix multiplication benchmark originally produced 43 kilobytes of trace data, but through our proposed static analysis techniques, LightningSimV2 reduces this to just 172 bytes---a \textbf{99.6\%} reduction. This greatly saves memory.

    \item Trace analysis time (\textbf{TA}) for optimized stall calculation using graph-based simulator. By avoiding event sorting and repeated event checking (discussed in Sec.~\ref{sec:LS-limitation}) in LightningSim, our graph-based simulation for stall calculation achieves up to \textbf{3.9\texttimes{}} speed up for trace analysis.

    \item Incremental stall calculation (\textbf{Incr.}) for efficient DSE. 
The most significant speedups, however, are in incremental simulation. LightningSimV2 achieves speedups by two orders of magnitude: up to \textbf{577\texttimes{}} with a geometric mean of \textbf{121\texttimes{}} speedup, trending higher for larger, more complex designs. 
This speedup implies that our decoupled graph construction and graph traversal is highly efficient, since the graph traversal step is extremely lightweight with superior scalability.

\end{itemize}

\subsection{Graph Optimization}

\begin{table}
    \caption{The effect of our proposed optimizations on the graph size.}
    \label{tab:graph-opt-results}
    \centering
    \setlength{\tabcolsep}{2pt}
    \renewcommand*{\arraystretch}{0.925}
    \footnotesize
    \begin{tabular}{l|c|c|c|c|c|c}
        \toprule
        & \multicolumn{2}{c|}{\textbf{Unoptimized}} & \multicolumn{2}{c|}{\textbf{Optimized}} & \multicolumn{2}{c}{\textbf{\% Reduced}} \\
        \textbf{Benchmark} & \textbf{Nodes} & \textbf{Edges} & \textbf{Nodes} & \textbf{Edges} & \textbf{Nodes} & \textbf{Edges} \\
        \midrule
        Pipelined nested loops~\cite{xilinx2021basic} & 408 & 407 & 2 & 1 & 99.51\% & 99.75\% \\
        AXI4 master~\cite{xilinx2021basic} & 176 & 181 & 7 & 12 & 96.02\% & 93.37\% \\
        Unoptimized FFT~\cite{kastner2018parallel} & 296K & 299K & 2.0K & 5.1K & 99.31\% & 98.29\% \\
        Multi-stage FFT~\cite{kastner2018parallel} & 3.9K & 3.9K & 10 & 20 & 99.74\% & 99.49\% \\
        Huffman encoding~\cite{kastner2018parallel} & 10K & 10K & 62 & 122 & 99.40\% & 98.83\% \\
        Matrix multiplication~\cite{kastner2018parallel} & 1.0K & 1.0K & 2 & 1 & 99.81\% & 99.90\% \\
        Parallelized merge sort~\cite{kastner2018parallel} & 138 & 141 & 6 & 9 & 95.65\% & 93.62\% \\
        Vector add with stream~\cite{xilinx2022vitis} & 17K & 41K & 16K & 41K & 1.35\% & 0.54\% \\
        FlowGNN GIN~\cite{flowgnn} & 236K & 419K & 56K & 238K & 76.37\% & 43.11\% \\
        FlowGNN GCN~\cite{flowgnn} & 143K & 323K & 70K & 250K & 51.05\% & 22.61\% \\
        FlowGNN GAT~\cite{flowgnn} & 150K & 227K & 7.4K & 84K & 95.09\% & 62.91\% \\
        FlowGNN PNA~\cite{flowgnn} & 392K & 501K & 34K & 143K & 91.32\% & 71.50\% \\
        FlowGNN DGN~\cite{flowgnn} & 136K & 257K & 43K & 164K & 68.75\% & 36.42\% \\
        \bottomrule
    \end{tabular}

\end{table}

We also analyze the impact of the optimizations discussed in Sec.~\ref{sec:node-elimination} that reduce the size of the graph by the node pending/commit technique and node/edge elimination. A representative subset is shown in Table~\ref{tab:graph-opt-results}.

We see that for most of the simple designs, our proposed graph optimization can significantly reduce the number of nodes and edges by more than 90\%. Even on more complex designs like the FlowGNN benchmarks, we typically see a reduction in nodes by more than 50\% and in edges by more than 20\%. 
Most designs have significant improvement, clearly demonstrating the effectiveness and memory efficiency of our optimizations.
Designs that are densely packed with events at nearly every stage do not benefit as much from our techniques, as demonstrated by the vector add with stream benchmark. 

\subsection{Larger Designs}\label{sec:larger-designs}

Sec.~\ref{sec:original-comparisons} only includes benchmarks where LightningSim's trace generation (TG) time dominates compared to its trace analysis (TA) time. To evaluate our scalability to larger designs, we test two designs where LightningSim's TA dominates TG. INR-Arch~\cite{inrarch} is a FIFO-heavy design for computing implicit neural representations for image processing applications. SkyNet~\cite{skynet} implements a DNN for object tracking.

The results in Table~\ref{tab:larger-designs} show LightningSimV2's scalability. Our graph-based approach reduces trace analysis time substantially, yielding up to \textbf{3.5\texttimes{}} overall speedup on large designs.

\begin{table}
    \caption{Results on larger designs.}
    \label{tab:larger-designs}
    \centering
    \begin{tabular}{l|r|r|r}
        \toprule
        \textbf{Benchmark} & \multicolumn{1}{c|}{\textbf{LS}} & \multicolumn{1}{c|}{\textbf{LSv2}} & \multicolumn{1}{c}{$\Delta$} \\
        \midrule
        \textbf{INR-Arch~\cite{inrarch}} & \textbf{7.56\,min.} & \textbf{2.14\,min.} & \textbf{3.5\texttimes} \\
        -- Trace generation & 1.50\,min. & 1.19\,min. & 1.3\texttimes{} \\
        -- Trace analysis & 6.05\,min. & 0.95\,min. & 6.4\texttimes{} \\
        -- Trace size & 2.88\,GB & 1.98\,GB & \textdownarrow\,31\% \\
        \midrule
        \textbf{SkyNet~\cite{skynet}} & \textbf{155.66\,min.} & \textbf{51.71\,min.} & \textbf{3.0\texttimes{}} \\
        -- Trace generation & 20.62\,min. & 16.70\,min. & 1.2\texttimes{} \\
        -- Trace analysis & 135.04\,min. & 35.01\,min. & 3.9\texttimes{} \\
        -- Trace size & 30.45\,GB & 29.92\,GB & \textdownarrow\,2\% \\
        \bottomrule
    \end{tabular}
\end{table}

\subsection{Memory Usage}

\begin{table}
    \caption{The effect of our proposed optimizations on RAM usage.}
    \label{tab:ram}
    \centering
    \begin{tabular}{l|rr|r}
        \toprule
        \textbf{Benchmark} & \multicolumn{1}{c}{\textbf{LS RAM}} & \multicolumn{1}{c|}{\textbf{LSv2 RAM}} & \multicolumn{1}{c}{\textbf{\% Reduced}} \\
        \midrule
        FlowGNN GIN~\cite{flowgnn} & 1.36\,GB & 1.03\,GB & \textminus24.62\% \\
        FlowGNN GCN~\cite{flowgnn} & 1.06\,GB & 0.91\,GB & \textminus14.22\% \\
        FlowGNN GAT~\cite{flowgnn} & 1.25\,GB & 0.91\,GB & \textminus26.69\% \\
        FlowGNN DGN~\cite{flowgnn} & 1.15\,GB & 0.92\,GB & \textminus19.40\% \\
        INR-Arch~\cite{inrarch} & 4.63\,GB & 2.15\,GB & \textminus53.56\% \\
        SkyNet~\cite{skynet} & 262.84\,GB & 59.35\,GB & \textminus77.42\% \\
        \bottomrule
    \end{tabular}
\end{table}

LightningSimV2 also boasts superior memory efficiency compared to LightningSim. We measure the maximum resident set size during execution and report the results in Table~\ref{tab:ram}. We see that our graph-based simulation techniques save up to \textbf{77\%} memory usage, especially on larger designs.

\section{Case Study: Fast DSE for FIFO Depths}
\label{sec:DSE}

In this case study, we demonstrate the parallelizability of our incremental stall calculation algorithm and see how it can be used for fast DSE for hundreds of FIFO depths.

\begin{figure}[t]
    \centering
    \includegraphics[width=\linewidth]{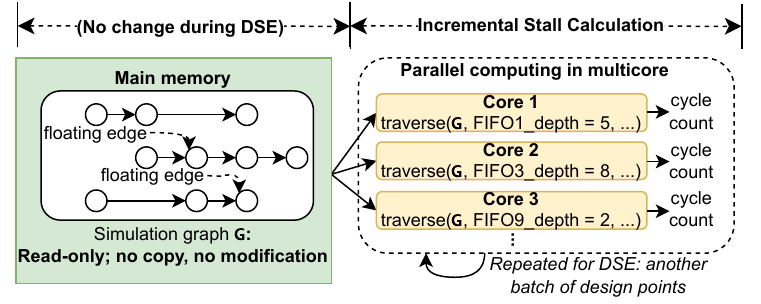}
    \caption{Parallel computing for DSE. The graph construction is a one-time effort: the compiled simulation graph $G$ will not be changed or modified, and thus is read-only and will not be copied to different cores. During DSE, only the incremental stall calculation step is re-executed in parallel with a batch of design points.}
    \label{fig:parallel-computing}
\end{figure}

In the original LightningSim, despite the nice feature of incremental stall calculation, it directly operates on and modifies event lists. This is unfriendly for parallel computing: to use multiple CPU cores, the event lists must be explicitly copied to each, incurring heavy communication and memory overhead.

In contrast, in LightningSimV2, the optimized graph structure can be trivially parallelized by running graph traversal across multiple CPU cores with different hardware parameters for floating edge resolution. This process is depicted in Fig.~\ref{fig:parallel-computing}.
We emphasize that the simulation graph \textit{does not} need to be copied to each core for parallel execution, since our graph traversal process does not mutate the graph itself, so it can be shared among all cores.
The only data that is kept specific to each core are the simulation hardware parameters and internal structures for computing clock cycles for each node.
This makes DSE easily scalable to large multiprocessor systems. 

To demonstrate the scalability of this process, we perform a case study using the INR-Arch~\cite{inrarch} design from Sec.~\ref{sec:larger-designs}, which is a large HLS dataflow design with 238 FIFO streams. After a full simulation of this design, we sample 128 points from the FIFO design space and process them in parallel on 64 CPU cores with both LightningSim and LightningSimV2.

\begin{figure}
    \centering
    \includegraphics[width=\linewidth]{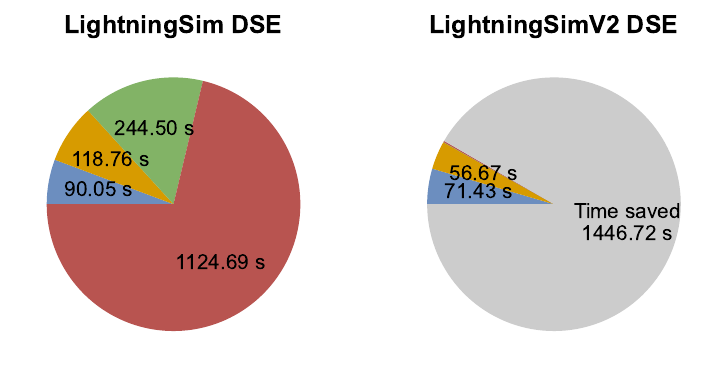}
    \caption{A breakdown of time spent when performing DSE for FIFO depths using LightningSim vs.\ LightningSimV2: blue represents trace generation, orange represents schedule resolution, green represents the time to evaluate a single design point, and red represents DSE across 128 design points. LightningSimV2 is so fast that the time needed to evaluate a single design point (0.53 seconds) and DSE across 128 design points (2.64 seconds) is nearly invisible in the chart.}
    \label{fig:dse-pie-chart}
\end{figure}

Fig.~\ref{fig:dse-pie-chart} shows the comparison. LightningSimV2 significantly speeds up every step of the simulation, clearly demonstrating its scalability and potential for DSE. Specifically, trace generation is 20\% faster thanks to our static analysis techniques, which reduced the trace size by 31\% from 2.88 GB to 1.98 GB. This reduction in trace data leads to 2.1\texttimes{} faster schedule resolution. 
The time it takes to evaluate a single design point reduces from over 4 minutes to 0.53 seconds. More importantly, the biggest benefit of LightningSimV2 is in the parallel evaluation of design points: what takes LightningSim over 18 minutes to perform using 64 cores is now completed by LightningSimV2 in just 2.64 seconds---a 426\texttimes{} speedup.

Overall, LightningSimV2 reduces the total experiment time by 24 minutes, or 91.7\%, demonstrating a clear advantage over LightningSim for DSE, especially for large-scale applications.

\section{Conclusion}

This work proposed LightningSimV2, a fast and scalable simulation tool for HLS designs. It addresses several inefficiencies in LightningSim through three major innovations:
fixed-loop bound optimization in IR trace generation, graph-based simulation using decoupled graph construction and graph traversal, and incremental parallelizable DSE.
Compared to the original LightningSim, LightningSimV2 achieves up to 3.5\texttimes{} speedup for end-to-end simulation, up to 6.4\texttimes{} speedup for trace analysis, and up to 577\texttimes{} speed up for incremental stall calculation.
We also demonstrated how these techniques make LightningSimV2 well-suited for efficient DSE of FIFO depth parameters even for designs with hundreds of FIFOs.

LightningSimV2 also offers new opportunities for future work. Although LightningSimV2, like the original LightningSim, does not yet support nondeterministic behavior in HLS designs, its simulation graph structure provides the groundwork for partial simulation of HLS designs, which could be useful in implementing nondeterministic constructs.

\bibliographystyle{IEEEtran}
\bibliography{reference}

\end{document}